\pdfoutput=1
\documentclass[a4paper,oneside,DIV=13,abstract=true]{scrartcl}
\usepackage{graphicx}
\usepackage{authblk}
\usepackage[square,numbers,sort&compress]{natbib}
\usepackage[USenglish]{babel}
\usepackage[separate-uncertainty = true, multi-part-units=single]{siunitx}
\usepackage[pdftex]{hyperref}                % manage links within the document. Needs to be the last loaded package.
\hypersetup{
	pdfauthor={Sylvain R. R. Iséni},
	pdftitle={Electric field vector mapping of guided ionization waves at atmospheric pressure},
	pdfsubject={report},
	pdfkeywords={.},
	pdfstartpage=1,
	bookmarksnumbered=true,
	bookmarksopen=true,
	colorlinks=true,
	citecolor=blue, %blue
	urlcolor=blue, %blue
	pdfstartview=FitH,
	pdfdisplaydoctitle=true,
	pdfpagelayout=OneColumn
}
\title{Electric field vector mapping of guided ionization waves at atmospheric pressure}
\author{Sylvain~Is\'{e}ni\thanks{sylvain.iseni@univ-orleans.fr --- \textsf{ORCiD:~0000-0002-4923-1657}}}
\affil{GREMI -- Groupe de Recherches sur l'\'{E}nerg\'{e}tique des Milieux Ionis\'{e}s\\
	UMR-7344, \textsc{cnrs}/Universit\'{e} d'Orl\'{e}ans, \textsc{France}.}
\begin{document}
\maketitle
\begin{abstract}
In this study, the dynamic of guided ionization wave (IW) generated by an atmospheric pressure plasma jet (APPJ) device in \emph{plasma gun} configuration is experimentally investigated.
The present work focuses on the properties of the intense electric field (EF) driving the IW.
Taking advantages of APPJs to produce \emph{guided} and \emph{reproducible} IWs, the induced EF vector is characterized spatially and temporally along the direction of propagation.
With this approach, EF vector mapping of guided IWs have been measured and documented for the first time.
In the first part, the propagation within a glass tube of the first IW is investigated.
Under the present conditions, a second guided IW is observed and propagates, leading to the formation of a guided streamer.
The EF due to transient charge deposited on the wall surface is observed, particularly at the end of the tube.
In the second part, one reports on the EF vector mapping under a dielectric substrate in contact with guided IWs.
EF strength up to \SI{55}{\kV\per\cm} has been measured and corroborates prior results from predictive numerical simulations.
Intriguing configurations of the EF lines will be of significant interest to validate theoretical models in order to refine the non-equilibrium plasma chemistry kinetics.
Furthermore, this preliminary work provides important insights into various applications involving IW driven discharges such as liquid activation, environmental treatments, plasma medicine, active flow control and plasma agriculture.
\end{abstract}
%%%%%%%%%%%%%% MANUSCRIPT %%%%%%%%%%%%%%
\section*{Introduction}
Room temperature atmospheric pressure plasma sources have been massively developed over the past decade aiming at various applications in the fields of biology and medicine~\citep{von_Woedtke_2013,Graves_2014}, environmental science and liquid activation~\citep{Foster_2012,Bruggeman_2016}, surface activation and material processing~\citep{Kim_2016,Penkov_2015} and active flow control~\citep{Moreau_2007}.
The uniqueness of non-thermal plasma combining different forms of energy --~electric, ionization, chemical, radiant and thermal~-- yields a rich reactive chemistry~\citep{Graves_2012}.
The production of reactive species depends strongly on the type of plasma source~\citep{Fridman2008}.
Among those, cold atmospheric pressure plasma jet (APPJ) offers the advantages of a cheap construction, a simple design and flexibility to integrate into industrial processes.
A large amount of electrical configurations and geometries have been reported in the literature~\citep{Lu_2012}.
Up-to-date additional technological aspects have been summarized in~\citep{Winter_2015}.\\
In most APPJ devices, the plasma formation is due to the propagation of guided ionization waves (IW) also named in the literature as pulsed atmospheric pressure plasma streams (PAPS)~\citep{Robert_2012} or more commonly \emph{guided streamer}~\citep{Boeuf2013}.
Operating at atmospheric pressure, atomic gases such as helium, argon or neon are used mainly for their chemical neutrality and ionization properties.
The morphology of guided IW depends on the gas and admixtures (typically N\textsubscript{2} or O\textsubscript{2}), producing a rather diffuse or filamentary discharge~\citep{Hofmann2012}.
The mechanism of guided IW has been extensively reviewed in~\citep{Lu_2014} from a theoretical and an experimental point of view.
Briefly, with the presence of high external electric field (EF), atoms and molecules of the medium are ionized by electron impacts leading to the formation of a space charge.
The density of volumetric charges contributes to the enhancement of the local EF in the front exceeding the external applied EF and allowing for the propagation of the IW.
Guided IW have the additional features of a predictable and reproducible direction propagation of the bright ionization front containing highly energetic electrons and photons being favorable for the creation of a darker --~partially~-- ionized channel.
The intense light emission of the guided IW front was observed first using high-speed photographs, where the term ``plasma bullet'' is suggested as a phenomenological description~\citep{Teschke_2005}.
Later, the evidence of the front is reported with the first measurement of EF strength up to \SI{26}{\kV\per\cm} in guided IW in helium, by means of Stark polarization emission spectroscopy~\citep{Sretenovi__2011} and confirms prior results of numerical simulations~\citep{Naidis_2010}.
Recent studies involving a similar plasma source have been carried out to characterize the EF resulting from the space charge ahead of the IW and before the first light coming from the ionization front~\citep{Robert_2015}.
In~\citep{Bourdon_2016}, the same group proposed a thoroughgoing numerical and experimental study where a successful agreement of the temporal evolution of the EF is found with the theoretical model.\\
Getting experimental insight into the EF resulting from guided IW is of high interest for the plasma community and beyond: towards large domains of applications.
Indeed, the electron temperature is strongly related to the EF and plays a key role in the ionization mechanisms as well as the global kinetic of the plasma chemistry.
Furthermore, the EF in guided IW is not uniform and is also time dependent which requires temporally and spatially resolved EF investigations also addressed as a future challenge in a recent review article~\citep{Lu_2014}.\\
In this paper, the authors report on an experimental study dealing with the characterization of the EF \emph{vector mapping} of a guided IW in helium at atmospheric pressure.
The work is centered around the simultaneous investigation of two orthogonal components of the EF space and time resolved.
The aspects documented here are twofold.
\begin{itemize}
	\item On the one hand, one focuses on the EF in the vicinity of the guided IW to prevent any perturbation induced by the measurement technique, and thus to study quantitatively the dynamic of the vector EF during the propagation.
	\item On the other hand, the EF properties of a guided IW in contact with a dielectric material will be presented to unveil the interaction between the plasma and a treated substrate.
\end{itemize}
\section*{Experimental setup}
In this work, guided IW are generated using a cold helium APPJ so-called \emph{plasma gun} (PG)~\citep{Robert_2012}.
\begin{figure}[hbtp]
	\centering
	\includegraphics[width=0.5\textwidth]{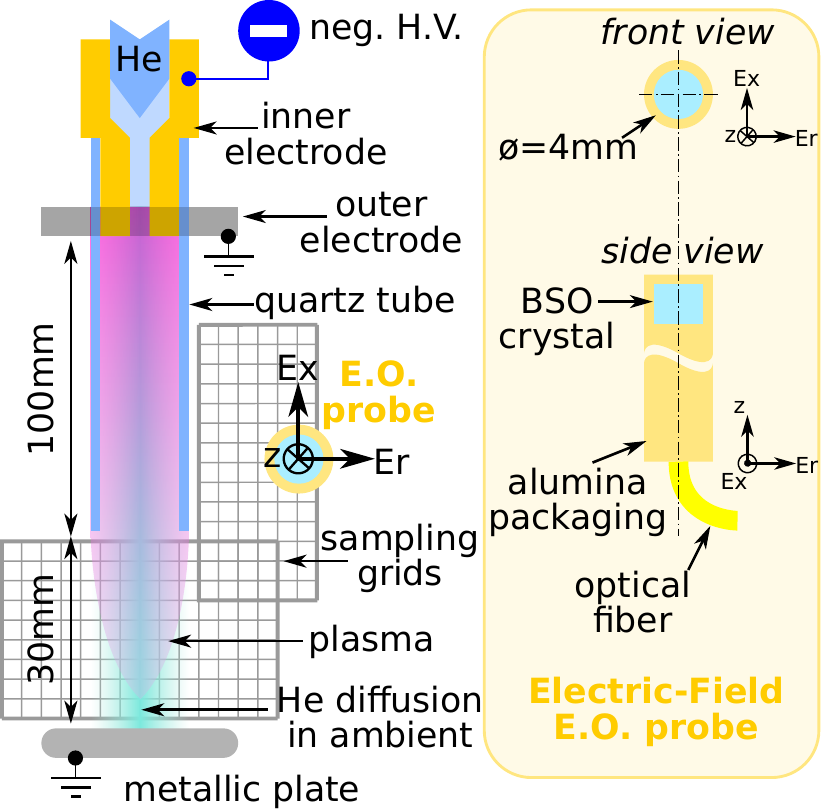}
	\caption{\label{fig:setup}Schematic of the experimental setup. On the left hand side, the plasma jet (PG) is depicted together with a conductive target and the measurement domains. On the right hand side, the E.O. probe used for the EF characterization is detailed with its two orthogonal axes of detection, $\vec{E_x}$ and $\vec{E_r}$.}
\end{figure}
In outlines, the PG consists of a coaxial dielectric barrier discharge design shown in figure~\ref{fig:setup}.
A concentric assembly involves a pair of electrodes mounted on a \SI{1.0}{\mm} thick and \SI{4.0}{\mm} inner diameter quartz tube.
The powered electrode fits into the tube and allows one to inject helium through a \SI{0.8}{\mm} channel drilled along the axis.
The grounded ring electrode is set on the outer surface of the tube and aligned with the tip of the powered electrode.
The helium flow rate is set at \SI{1.0}{\liter\per\min} at standard conditions of temperature and pressure.
The inner electrode is connected to a symmetric H.V. pulsed generator.
In this work, repetitive negative pulses of \SI{2.4}{\us} FWHM and \SI{1.3}{\us} fall time at the frequency of \SI{1.0}{\kHz} are used exclusively.
The H.V. negative waveform is measured on the electrode and plotted in figure~\ref{fig:voltage}.
The choice of the negative polarity is twofold: from the emission of the light, the plasma seems homogeneous in volume and the gas flow pattern exiting the tube remains laminar in these conditions~\citep{Robert_2015,Darny_2016}.
This contributes to the stability of the rectilinear direction of the guided IW which propagates into the helium channel expanding in the ambient air.
The partially ionized gas outside the quartz tube is the so-called \emph{plasma plume}.
Most of the geometries of APPJs cannot provide a fixed electric potential in the vicinity of the plasma plume which makes APPJs very sensitive to the electrical composition in its surrounding.
Thus, the substrate of a treated sample is an important element of the electrical circuit of APPJs.
Therefore, a conductive plate connected to the ground is placed \SI{30}{\mm} away from the tube exit to fix a known electric potential in the vicinity of the plasma plume.
It is important to mentioned that the guided IW --~and thus the plasma plume~-- does not reach its surface, preventing a direct electric connection.\\
The EF diagnostic is carried out with a customized pigtailed electro-optic sensor (E.O.) commercialized by \textsc{Kapteos~s.a.s.} and detailed in figure~\ref{fig:setup}.
This technology applied recently to plasmas~\citep{GaboritReuterIseniEtAl2014} is based on the Pockels effect.
The sensor is made of a cylindrical isotropic birefringent crystal probed with a laser beam guided through a fiber optic~\citep{Gaborit_2014}.
Depending on the variation of the light polarization induced by the external EF, the system is able to measure simultaneously the EF amplitude in two orthogonal directions of space.
Although this method is not suitable for the measurement of static EF, the instrument is able to detect transient EF from tens of \si{kHz} up to \SI{1}{\GHz} with a time resolution of \SI{1}{\ns} and a spatial resolution of the crystal size (\SI{1.8}{\mm}$\times$\SI{2.0}{\mm}).\\
To perform a 2D characterization of the EF, the position in space the E.O. sensor is adjusted with a \mbox{X-Y}~motorized translation stage.
As shown in figure~\ref{fig:setup}, $\vec{E_x}$ axis of the E.O. sensor is parallel to the longitudinal axis of the PG while orthogonal axis, $\vec{E_r}$, is collinear with the radius of the tube.
Two sampling domains are defined: one on the right side of the PG --~along the tube~-- in order to characterize the EF of the guided IW without any disturbance from the probe.
A second domain is in the gap between the exit of the tube and the metallic plate.
In the later condition, the E.O. probe is directly in contact with the plasma plume and acts as a dielectric treated substrate.
In each node of the sampling grid, the time evolution of longitudinal (E\textsubscript{x}) and radial (E\textsubscript{r}) EF is measured simultaneously.
Hence, one defines the EF strength by $\left\| \vec{E} \right\| = \left(E_x^2 + E_r^2 \right)^{\frac{1}{2}}$.
\section*{Results and discussion}
\begin{figure}[!ht]
	\centering
	\includegraphics[width=0.66\textwidth]{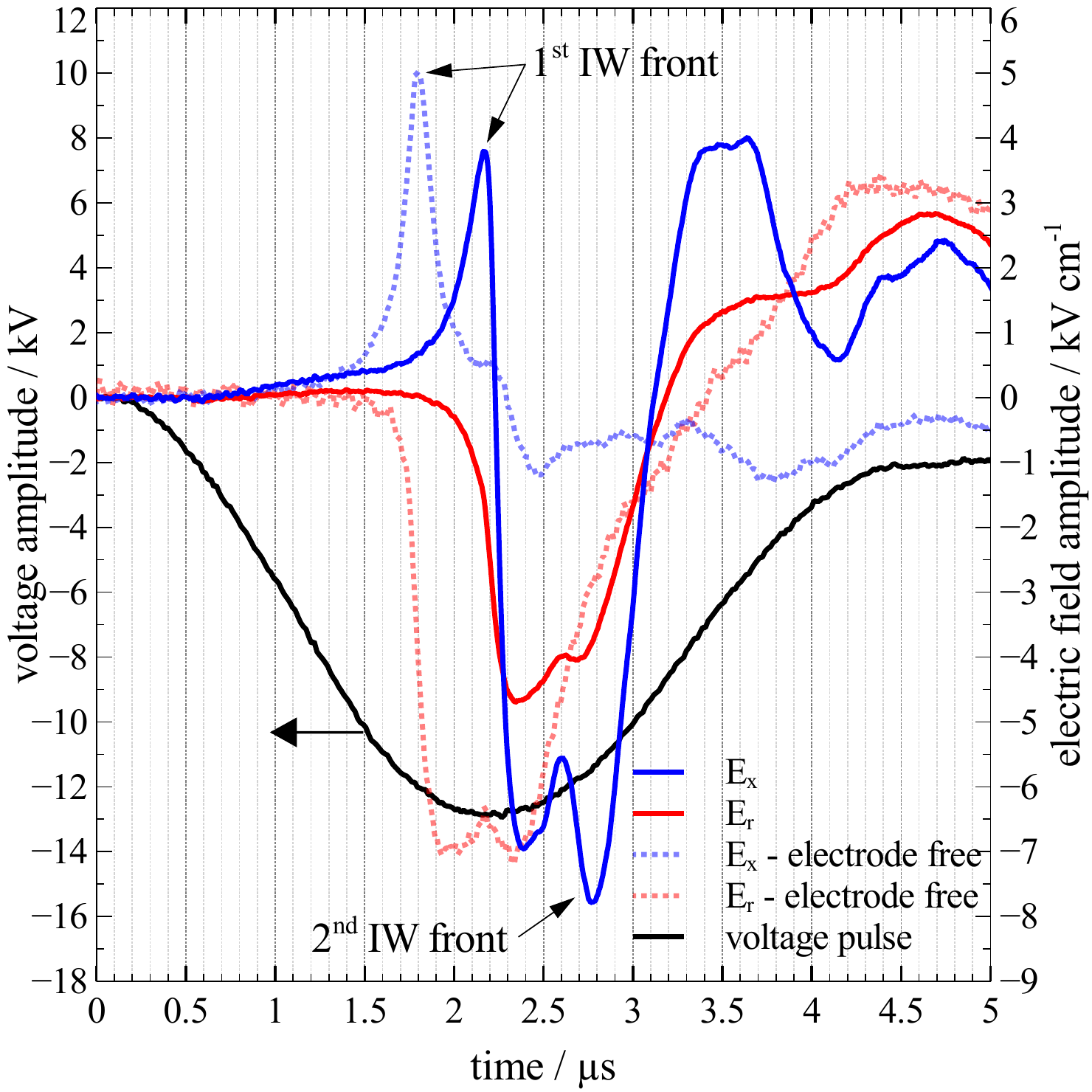}
	\caption{\label{fig:voltage} Time evolution of the voltage pulse, the axial (E\textsubscript{x}) and radial (E\textsubscript{r}) amplitude of the EF. The center of the E.O. probe is located at \SI{5.2}{\mm} from the PG axis. The term ``electrode free'' is used when the PG operates without any grounded ring electrode like in~\citep{Bourdon_2016}. In the latter, the measurement is performed \SI{50}{\mm} from the tip of the inner electrode while in the case of continuous lines, the signals were acquired \SI{25}{\mm} further.}
\end{figure}

\begin{figure*}[!ht]
	\centering	
	\includegraphics[width=\textwidth]{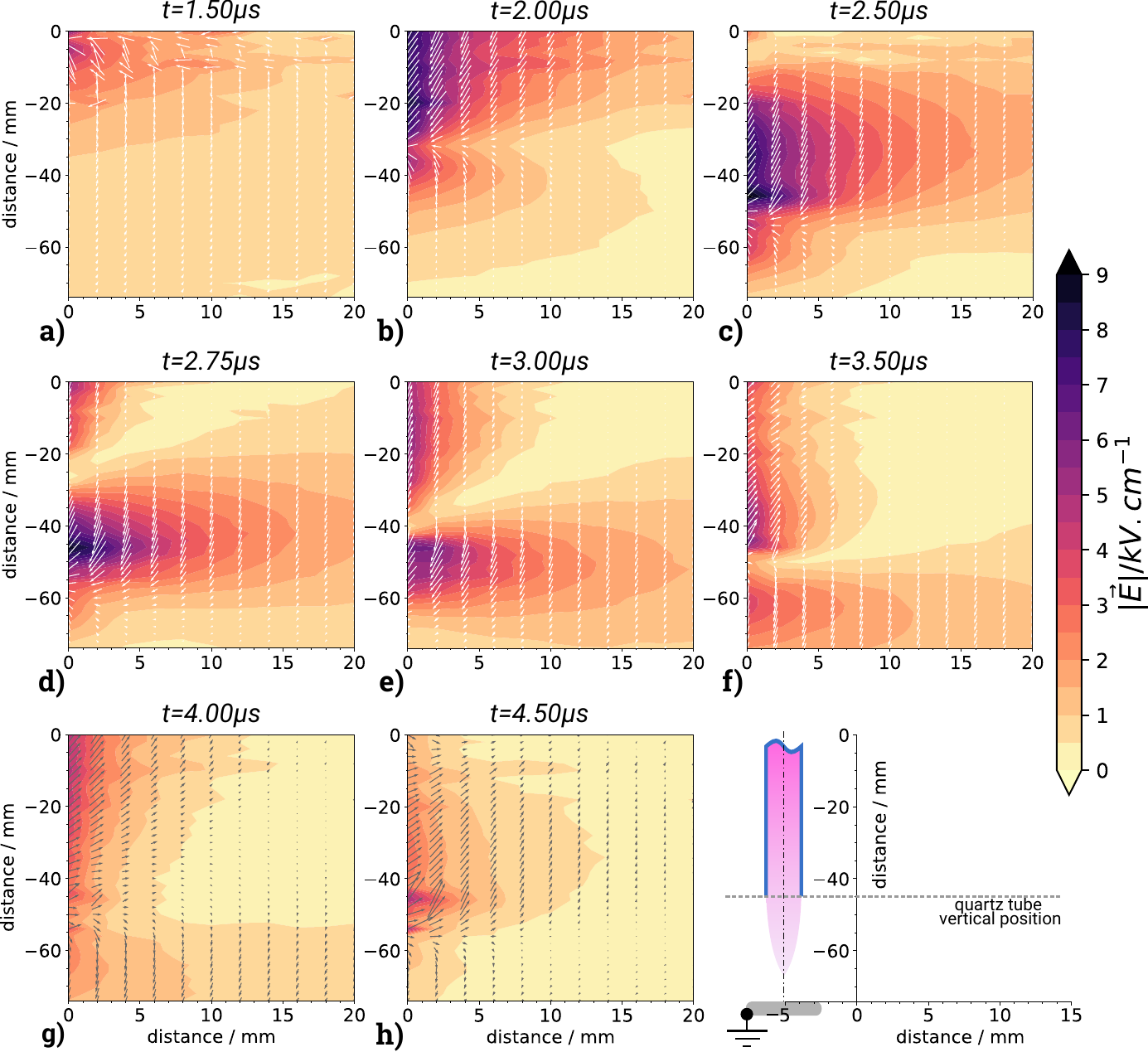}
	\caption{\label{fig:seq_IW_tube} Sequence of EF vector mapping on the right hand side of the propagation of the guided IW from the PG. On the vertical axis, the origin is located \SI{55}{\mm} from the tip of the inner electrode so that \SI{-75}{\mm} corresponds to the surface of the grounded metallic plate (see figure~\ref{fig:setup}). Thus, the tip of the quartz tube is at \SI{-45}{\mm}. The guided IW propagation being from top to bottom, the distance values are negative by convention. On the horizontal axis, the origin --~being the closest to the tube~-- corresponds to the center of the E.O. probe crystal positioned \SI{5.2}{\mm} from the PG axis.}
\end{figure*}

An example of the time evolution of both E\textsubscript{x} and E\textsubscript{r} are presented in figure~\ref{fig:voltage}.
The set of curves in continuous line are related to this study while the set drawn in dash line refers to prior investigations correlating successfully experimental and numerical simulations results~\citep{Bourdon_2016}.
Focusing first on E\textsubscript{x} signals, the front of the IW is clearly identified as a sharp peak of about \SI{200}{\ns} with an amplitude of several \si{\kV\per\cm} as already reported in~\citep{Robert_2015,Bourdon_2016,Darny_2016}.
A time delay between both signals is found around \SI{360}{\ns} due to the two different vertical positions of the measurement.
The propagation velocity of the IW front is estimated around \SI{7E6}{\cm\per\s} which is in agreement with prior studies~\citep{Bourdon_2016}.
Shortly after the IW front passed, one notices a quick rise of E\textsubscript{r} resembling a square pulse of \SI{1}{\us} FWHM with an amplitude between \SI{4.5}{\kV\per\cm} and \SI{8}{\kV\per\cm}.
The reason of this intense radial EF is explained by the creation of a --~partially~-- conductive channel formed within the tube after the propagation of the IW front.
Due to a significant density of charges and the direct contact of the plasma channel with the powered electrode, the square feature in the E\textsubscript{r} signals result from the dynamic of the H.V. pulse.
The electric potential is thus transferred from the inner electrode to the plasma column from where results the intense radial EF.
Depending on the electrical configuration of the PG, the dynamic of the voltage pulse can have a significant influence on E\textsubscript{x}~\citep{Darny_2016}.
Indeed, in case of a PG free of grounded ring electrode, one can observe a negative peak occurring at \SI{2.4}{\us} which corresponds exactly on time with the applied voltage imposed by the generator and rising to the ground potential.
This negative peak of E\textsubscript{x} is drastically enhanced (down to \SI{-7}{\kV\per\cm}) for a PG reactor equipped with a grounded electrode.
Furthermore, on the same signal, a global minimum of E\textsubscript{x} is seen at \SI{2.8}{\us} \textit{i.e.} \SI{0.6}{\us} after the IW front.
The latter results from the second IW front propagation as discussed in~\citep{Darny_2017}.\\
Figure~\ref{fig:seq_IW_tube} presents a sequence of EF vector mapping sampled on the right hand side of a PG.
The data shown in those maps are time dependent signals of E\textsubscript{x} and E\textsubscript{r} --~such as shown in figure~\ref{fig:voltage}~-- recorded for different nodes of the sampling grid.
By processing the data set, the vector field can be displayed, revealing the EF lines together with the EF strength.
The direct visualization of the IW front is depicted in figure~\ref{fig:seq_IW_tube}a) with a strong EF gradient and the vectors pointing toward the front.
The direction of the EF results from the negative charge --~mainly electron due to their higher mobility~-- ahead of the IW.
At \SI{2.2}{\us}, the H.V. pulse reaches its minimum at \SI{-13}{\kV} imposed to the plasma channel.
Figure~\ref{fig:seq_IW_tube}b) shows the EF resulting from the negative charge attached to the inner surface of the tube balancing the positive space charge which bounds the plasma column~\citep{Norberg_2015}.
This is sustained as long as the level of the applied voltage is not back to the ground potential.
Thus an intense EF from the plasma column extend up to \SI{-40}{\mm} and giving rise to a \emph{negative (anode-directed) guided streamer}~\citep{9783642647604}.
The propagation of the IW starts to slow down around \SI{2.4}{\us}, \textit{i.e} when the electric potential is at the minimum.
Then, the latter begins to raise up, leading to a fast increase of the electric potential of the inner electrode which stops the propagation of the first IW as shown in figure~\ref{fig:seq_IW_tube}c).
The EF strength drops quickly as well as the electron kinetics which will affects the plasma properties, for instance He metastable densities~\citep{Darny_2017}.
The localized intense EF at \SI{-45}{\mm} corresponds to the end of the tube and indicates a significant attachment of transient charge distributed around the tube section.
Intuitively, the magnitude of the induced EF is dependent on the shaping of the tube edge and the density of charge.
On figure~\ref{fig:seq_IW_tube}d), one notices an enhancement of charge and the resulting EF measured over several tens of \si{\ns}.
The direction of the EF indicates the presence of negative charge.
Back diffusion of ambient air into the He gas flow was already observed, particularly with the emission of excited atomic oxygen~\citep{Iseni2016a,Iseni2016}.
According to~\citep{Norberg_2015}, O\textsubscript{2}\textsuperscript{$-$} are the dominant negative ions at the exit of the tube due to a smaller T\textsubscript{e} being favorable to electrons to attach to O\textsubscript{2}.
Although the technique involved in this study cannot provide any information about static charge, deposited charges have been recently investigated~\citep{Wild_2013,Slikboer_2017}.
In~\citep{Slikboer_2017}, the authors estimated the loss within the tube around \SI{0.1}{\nano\coulomb\per\square\cm}.\\
The fast rise of the electrode potential induces a second IW front propagating with a strong EF polarized in the opposite direction with the first IW.
The separation between both IW is characterized by a sharp gradient of EF creating a gap where the value of the EF strength is damped close to \SI{0}{\kV\per\cm}.
At \SI{2.75}{\us}, the applied potential imposed by the generator is distributed along the conductive channel after the propagation of the second IW.
Consequently, the EF direction reverses due to an electrode potential being increased compared to the charge residual of the first IW and the charging of the wall tube by positive charge~\citep{Xiong_2012}.
This brutal flipping of the EF influences the ionization dynamic and the electron kinetic which govern the whole plasma chemistry kinetic.
The propagation of the second IW within the tube is shown in figure~\ref{fig:seq_IW_tube}e).
The contribution of the negative charge at the tube exit to weaken significantly the EF of the second IW is clearly identified.
As the result, the shaping of the tube exit must play a key role in the dynamic of the IW propagation and the plasma dynamics due to the presence of a local EF of tens of \si{\kV\per\cm}.
The plasma column is sustained until the applied voltage returns back to the ground potential.
The positive rise induces the propagation of the second IW leading to the formation of a \emph{positive (cathode-directed) guided streamer}~\citep{9783642647604}.
Indeed, the direction of the EF vectors of the plasma channel indicates a positive charging of the tube wall and confirms prior predictive numerical work regarding positive guided streamer mechanism~\citep{Xiong_2012}.
Last but not least, the EF strength is measured to be higher in the column of the negative streamer compared to the positive streamer which is in agreement with several predictive models~\citep{Naidis_2011,Xiong_2013}.\\
APPJs are also driven with AC H.V. power supply where the existence of a second IW front may be questioned.
According to~\citep{Sretenovic2014}, with slow AC regime, a IW is ignited at each half-period.
Further investigations are necessary to identify the role of the frequency and the H.V. amplitude for the generation of a second guided IW.\\
The second part of this study focuses on the EF vectors in the plasma plume region.
In contrast with the previous experiment, the E.O. probe is now being directly in contact with the IW during scanning operations.
\begin{figure*}[!ht]
	\centering
	\includegraphics[width=\textwidth]{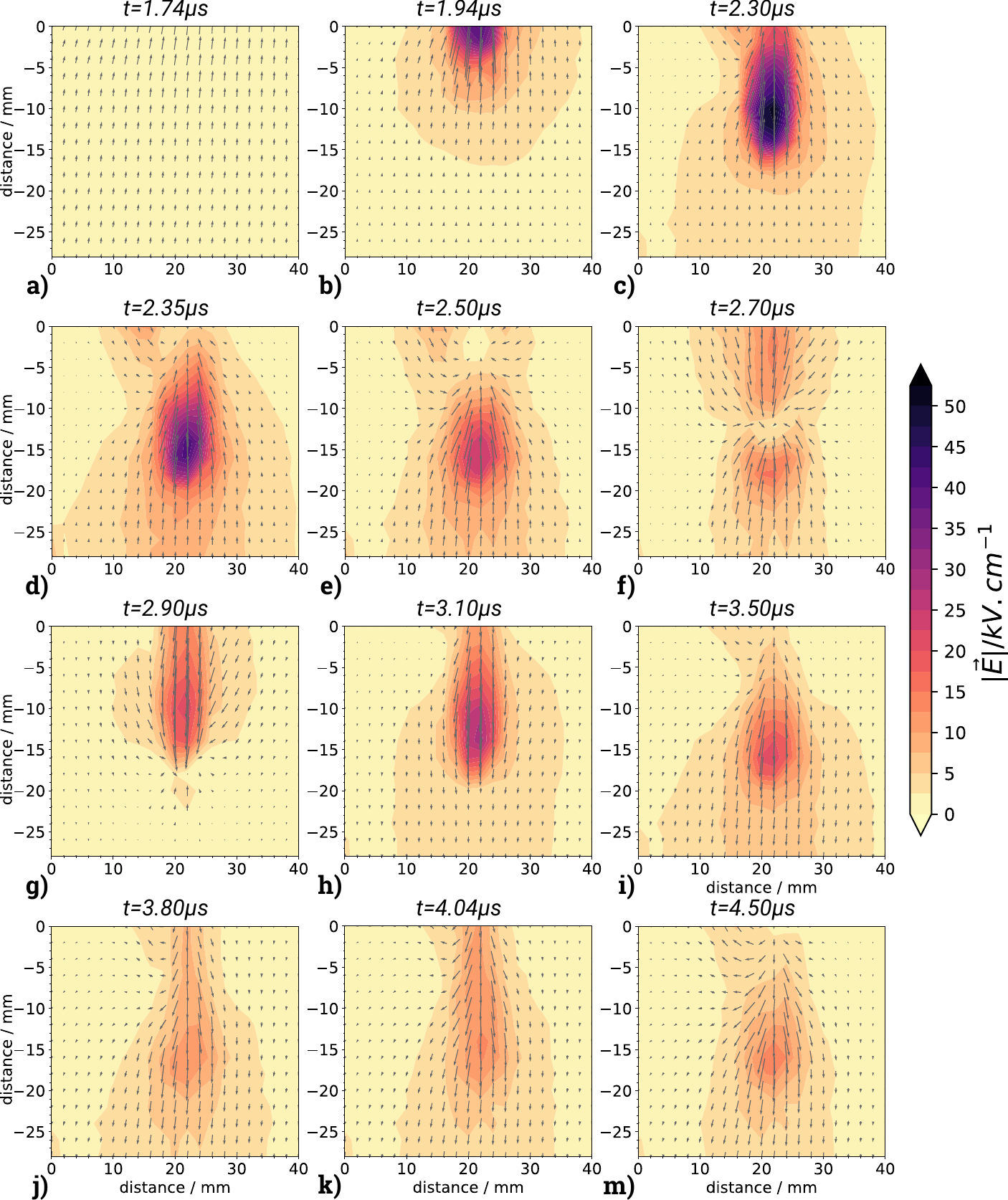}
	\caption{\label{fig:seq_IW}Sequence of EF vector mapping at the PG exit. On the vertical axis, the origin corresponds to the tube end \textit{i.e.} \SI{100}{\mm} from the inner electrode. The position of the surface of the grounded metallic plate is at \SI{-30}{\mm} (see figure~\ref{fig:setup}). The guided IW propagation being from top to bottom, the distance values are negative by convention. On the horizontal axis, the axis of the PG tube is at \SI{22}{\mm}.}
\end{figure*}
Sequences of EF vectors maps are presented in figure~\ref{fig:seq_IW}.
Depending on the position of the E.O. probe, the plasma plume can impinge on the alumina surface protecting the BSO crystal.
In this case, it is important to clarify that the interpretation of the results differs from the previous set.
Indeed, the EF strength values shown in figure~\ref{fig:seq_IW} are not within the IW front, but correspond to the EF intensity under a dielectric surface --~of a relative permittivity ($\varepsilon_r = \varepsilon/\varepsilon_0$), $\varepsilon \approx 10$~-- touching the plasma plume.
However, the directions of the EF vectors are assumed to do not be affected by the surface.
In figure~\ref{fig:seq_IW}a), the EF vectors is polarized vertically, directed toward the negative charge ahead of the IW front which is propagating into the tube.
At \SI{1.94}{\us}, the IW front has already exited the tube and propagates into the ambient.
An intense EF expands within a symmetric volume of \SI{30}{\mm} large and a maximum strength up to \SI{50}{\kV\per\cm}.
Such extreme EF measured under the \SI{1.0}{\mm} dielectric layer results from the interaction of the plasma with the surface.
In the present case, the relatively low $\varepsilon_r$, the continuity of the electric displacement vector, $\vec{D}=\varepsilon \vec{E}$ allows for an effective penetration of the longitudinal EF of the IW front through the surface.
The EF strength is enhanced due to the transient charge deposited on top of the surface.
A recent numerical work on plasma jet interacting with various dielectric and metal surfaces documents EF strength value in good agreement with the present experiment~\citep{Norberg_2015}.
The computed value under the surface of $\varepsilon_r = 10$ ranges from \SIrange{14}{40}{\kV\per\cm}.
Although in the current study, it is not possible to untangle the contribution of the IW front and the transient charge to the EF strength, the maximum value is reached around \SI{2.30}{\us}, when the first IW front exits the tube.
In figure~\ref{fig:seq_IW}c) and d), the applied voltage pulse has passed its minimum amplitude to start the rising slope which leads to a separation of the first IW as previously mentioned.
Interestingly, the EF direction in the vicinity of the tube exit describes a curl which is attributed to the presence of negative charge, such as O\textsubscript{2}\textsuperscript{$-$} at the tip of the glass tube~\citep{Norberg_2015}.\\
The rise up of the applied voltage imposed to the plasma channel implies a flip of the EF and the creation of a second IW shown in figure~\ref{fig:seq_IW}e).
At this time (\SI{2.4}{\us}), an isolated EF source up to \SI{25}{\kV\per\cm} is observed at \SI{-15}{\mm} being generated by negative charge stored on the dielectric surface.
The recombination of the charges leads to a significant decrease of the EF strength while the second IW propagates.
The dynamic of both processes gives rise to an intriguing configuration of the EF lines as shown in figure~\ref{fig:seq_IW}f).
The map reveals a head-to-tail EF configuration separated by a gap at \SI{-12}{\mm} where the EF induced by the second IW and the surface charge compensate each other.
It is anticipated that such EF feature must have a significant influence on the ionization mechanism and chemical kinetic of the plasma.
The recombination of the charge on top of the surface by the second IW allows the latter to propagate further.
At \SI{2.9}{\us}, the EF is polarized vertically, flipped upside down, revealing the path of the ionized channel produced by the IW propagation.
The structure of the EF and the magnitude up to \SI{30}{\kV\per\cm} evidences the existence of a second IW propagating in to the mixture of helium and air.
The charging of the surface significantly contributes to the EF strength after the applied voltage returned back to the ground potential.
Knowing the key role of the EF the production mechanism of reactive species, it is assumed that the whole plasma chemistry must be strongly affected in case of liquid substrate in contact with IW.\\
% CONCLUSION
\section*{Concluding remarks}
Focusing on the essence of APPJs --guided IWs--, this manuscript reports for the first time on experimental EF vector characterization.
Measurements performed with a unique technology based on E.O. probe allow us to produce EF vector mapping of guided IWs.
Propagation velocities of the IW are in agreement with prior works based on the investigation of light emission. 
It is shown that APPJs powered with a \si{\us} H.V. pulse generator initiate the propagation of two consecutive guided IW with opposite EF vector direction.
It is discussed that the time scale of the H.V. pulse and the remaining ionization of the channel created by the first guided IW are crucial parameters for the generation of a second guided IW.
The transient EF due to the attachment of charge on the wall has been identified and a specific attention must be paid to the shaping of the section of the tube.\\
EF vector mapping measured under a dielectric surface in contact with the IW brings new insights into the EF lines complexity driving the electron kinetic and therefore the global non-equilibrium chemistry.
A maximum of \SI{55}{\kV\per\cm} is measured under the dielectric surface.
The dynamic of the first IW is consistent non-intrusive measurements although the presence of a dielectric substrate in contact with the IW implies a change of the electric properties.
The evidence of a second guided IW is documented and unveils an intriguing head-to-tail configuration of the EF vector direction.
This preliminary study opens new ways to investigate experimentally EF induced by guided IWs and derived plasma devices such as APPJs.	
A plethora of applications will benefit from EF vector characterization to understand and to optimize plasma sources and processes, particularly in the field of plasma liquid interaction, surface modification, plasma medicine and agriculture including plasma cell interactions.

% ACKNOWLEDGEMENTS
\section*{Acknowledgment}
The author would like to thank X.~Damany, J.~M.~Pouvesle and E.~Robert for the continuous encouragement and general support, making this work possible.
S.~Dozias is acknowledged for the electronic engineering and technical support.\\
Data and graphs have been exclusively processed with open-source libraries SciPy, NumPy~\citep{van_der_Walt_2011}, Veusz and Matplotlib~\citep{Hunter_2007}.
% REFERENCES
\bibliographystyle{unsrtnat}
\bibliography{2D_EF_GIW_APPJ_IS_2017}

\begin{thebibliography}{36}
\providecommand{\natexlab}[1]{#1}
\providecommand{\url}[1]{\texttt{#1}}
\expandafter\ifx\csname urlstyle\endcsname\relax
  \providecommand{\doi}[1]{doi: #1}\else
  \providecommand{\doi}{doi: \begingroup \urlstyle{rm}\Url}\fi

\bibitem[von Woedtke et~al.(2013)von Woedtke, Reuter, Masur, and
  Weltmann]{von_Woedtke_2013}
Th. von Woedtke, S.~Reuter, K.~Masur, and K.-D. Weltmann.
\newblock Plasmas for medicine.
\newblock \emph{Physics Reports}, 530\penalty0 (4):\penalty0 291--320, sep
  2013.
\newblock \doi{10.1016/j.physrep.2013.05.005}.
\newblock URL \url{http://dx.doi.org/10.1016/j.physrep.2013.05.005}.

\bibitem[Graves(2014)]{Graves_2014}
David~B. Graves.
\newblock Low temperature plasma biomedicine: A tutorial review.
\newblock \emph{Physics of Plasmas}, 21\penalty0 (8):\penalty0 080901, aug
  2014.
\newblock \doi{10.1063/1.4892534}.
\newblock URL \url{http://dx.doi.org/10.1063/1.4892534}.

\bibitem[Foster et~al.(2012)Foster, Sommers, Gucker, Blankson, and
  Adamovsky]{Foster_2012}
John Foster, Bradley~S. Sommers, Sarah~Nowak Gucker, Isaiah~M. Blankson, and
  Grigory Adamovsky.
\newblock Perspectives on the interaction of plasmas with liquid water for
  water purification.
\newblock \emph{{IEEE} Transactions on Plasma Science}, 40\penalty0
  (5):\penalty0 1311--1323, may 2012.
\newblock \doi{10.1109/tps.2011.2180028}.
\newblock URL \url{http://dx.doi.org/10.1109/tps.2011.2180028}.

\bibitem[Bruggeman et~al.(2016)Bruggeman, Kushner, Locke, Gardeniers, Graham,
  Graves, Hofman-Caris, Maric, Reid, Ceriani, Rivas, Foster, Garrick, Gorbanev,
  Hamaguchi, Iza, Jablonowski, Klimova, Kolb, Krcma, Lukes, Machala, Marinov,
  Mariotti, Thagard, Minakata, Neyts, Pawlat, Petrovic, Pflieger, Reuter,
  Schram, Schröter, Shiraiwa, Tarabov{\'{a}}, Tsai, Verlet, von Woedtke,
  Wilson, Yasui, and Zvereva]{Bruggeman_2016}
P~J Bruggeman, M~J Kushner, B~R Locke, J~G~E Gardeniers, W~G Graham, D~B
  Graves, R~C H~M Hofman-Caris, D~Maric, J~P Reid, E~Ceriani, D~Fernandez
  Rivas, J~E Foster, S~C Garrick, Y~Gorbanev, S~Hamaguchi, F~Iza,
  H~Jablonowski, E~Klimova, J~Kolb, F~Krcma, P~Lukes, Z~Machala, I~Marinov,
  D~Mariotti, S~Mededovic Thagard, D~Minakata, E~C Neyts, J~Pawlat, Z~Lj
  Petrovic, R~Pflieger, S~Reuter, D~C Schram, S~Schröter, M~Shiraiwa,
  B~Tarabov{\'{a}}, P~A Tsai, J~R~R Verlet, T~von Woedtke, K~R Wilson, K~Yasui,
  and G~Zvereva.
\newblock Plasma{\textendash}liquid interactions: a review and roadmap.
\newblock \emph{Plasma Sources Science and Technology}, 25\penalty0
  (5):\penalty0 053002, sep 2016.
\newblock \doi{10.1088/0963-0252/25/5/053002}.
\newblock URL \url{http://dx.doi.org/10.1088/0963-0252/25/5/053002}.

\bibitem[Kim et~al.(2016)Kim, Lee, Mishra, and Yeom]{Kim_2016}
Kyong~Nam Kim, Seung~Min Lee, Anurag Mishra, and Geun~Young Yeom.
\newblock Atmospheric pressure plasmas for surface modification of flexible and
  printed electronic devices: A review.
\newblock \emph{Thin Solid Films}, 598:\penalty0 315--334, jan 2016.
\newblock \doi{10.1016/j.tsf.2015.05.035}.
\newblock URL \url{http://dx.doi.org/10.1016/j.tsf.2015.05.035}.

\bibitem[Penkov et~al.(2015)Penkov, Khadem, Lim, and Kim]{Penkov_2015}
Oleksiy~V. Penkov, Mahdi Khadem, Won-Suk Lim, and Dae-Eun Kim.
\newblock A review of recent applications of atmospheric pressure plasma jets
  for materials processing.
\newblock \emph{Journal of Coatings Technology and Research}, 12\penalty0
  (2):\penalty0 225--235, jan 2015.
\newblock \doi{10.1007/s11998-014-9638-z}.
\newblock URL \url{http://dx.doi.org/10.1007/s11998-014-9638-z}.

\bibitem[Moreau(2007)]{Moreau_2007}
Eric Moreau.
\newblock Airflow control by non-thermal plasma actuators.
\newblock \emph{Journal of Physics D: Applied Physics}, 40\penalty0
  (3):\penalty0 605--636, jan 2007.
\newblock \doi{10.1088/0022-3727/40/3/s01}.
\newblock URL \url{http://dx.doi.org/10.1088/0022-3727/40/3/s01}.

\bibitem[Graves(2012)]{Graves_2012}
David~B Graves.
\newblock The emerging role of reactive oxygen and nitrogen species in redox
  biology and some implications for plasma applications to medicine and
  biology.
\newblock \emph{Journal of Physics D: Applied Physics}, 45\penalty0
  (26):\penalty0 263001, jun 2012.
\newblock \doi{10.1088/0022-3727/45/26/263001}.
\newblock URL \url{http://dx.doi.org/10.1088/0022-3727/45/26/263001}.

\bibitem[Fridman(2008)]{Fridman2008}
Alexander Fridman.
\newblock \emph{Plasma Chemistry}.
\newblock Cambridge University Press, July 2008.
\newblock ISBN 9780511546075.

\bibitem[Lu et~al.(2012)Lu, Laroussi, and Puech]{Lu_2012}
X~Lu, M~Laroussi, and V~Puech.
\newblock On atmospheric-pressure non-equilibrium plasma jets and plasma
  bullets.
\newblock \emph{Plasma Sources Science and Technology}, 21\penalty0
  (3):\penalty0 034005, apr 2012.
\newblock \doi{10.1088/0963-0252/21/3/034005}.
\newblock URL \url{http://dx.doi.org/10.1088/0963-0252/21/3/034005}.

\bibitem[Winter et~al.(2015)Winter, Brandenburg, and Weltmann]{Winter_2015}
J~Winter, R~Brandenburg, and K-D Weltmann.
\newblock Atmospheric pressure plasma jets: an overview of devices and new
  directions.
\newblock \emph{Plasma Sources Science and Technology}, 24\penalty0
  (6):\penalty0 064001, oct 2015.
\newblock \doi{10.1088/0963-0252/24/6/064001}.
\newblock URL \url{http://dx.doi.org/10.1088/0963-0252/24/6/064001}.

\bibitem[Robert et~al.(2012)Robert, Sarron, Ri{\`{e}}s, Dozias, Vandamme, and
  Pouvesle]{Robert_2012}
E~Robert, V~Sarron, D~Ri{\`{e}}s, S~Dozias, M~Vandamme, and J-M Pouvesle.
\newblock Characterization of pulsed atmospheric-pressure plasma streams
  ({PAPS}) generated by a plasma gun.
\newblock \emph{Plasma Sources Science and Technology}, 21\penalty0
  (3):\penalty0 034017, may 2012.
\newblock \doi{10.1088/0963-0252/21/3/034017}.
\newblock URL \url{http://dx.doi.org/10.1088/0963-0252/21/3/034017}.

\bibitem[Boeuf et~al.(2013)Boeuf, Yang, and Pitchford]{Boeuf2013}
J-P Boeuf, L~L Yang, and L~C Pitchford.
\newblock {Dynamics of a guided streamer (``plasma bullet'') in a helium jet in
  air at atmospheric pressure}.
\newblock \emph{{Journal of Physics~D: Applied Physics}}, 46\penalty0
  (1):\penalty0 015201, jan 2013.
\newblock \doi{10.1088/0022-3727/46/1/015201}.
\newblock URL \url{http://dx.doi.org/10.1088/0022-3727/46/1/015201}.

\bibitem[Hofmann et~al.(2012)Hofmann, Sobota, and Bruggeman]{Hofmann2012}
S.~Hofmann, A.~Sobota, and P.~Bruggeman.
\newblock Transitions between and control of guided and branching streamers in
  {DC} nanosecond pulsed excited plasma jets.
\newblock \emph{{IEEE} Transactions on Plasma Science}, 40\penalty0
  (11):\penalty0 2888--2899, Nov 2012.
\newblock \doi{10.1109/TPS.2012.2211621}.
\newblock URL \url{http://dx.doi.org/10.1109/TPS.2012.2211621}.

\bibitem[Lu et~al.(2014)Lu, Naidis, Laroussi, and Ostrikov]{Lu_2014}
X.~Lu, G.V. Naidis, M.~Laroussi, and K.~Ostrikov.
\newblock Guided ionization waves: Theory and experiments.
\newblock \emph{Physics Reports}, 540\penalty0 (3):\penalty0 123--166, jul
  2014.
\newblock \doi{10.1016/j.physrep.2014.02.006}.
\newblock URL \url{http://dx.doi.org/10.1016/j.physrep.2014.02.006}.

\bibitem[Teschke et~al.(2005)Teschke, Kedzierski, Finantu-Dinu, Korzec, and
  Engemann]{Teschke_2005}
M.~Teschke, J.~Kedzierski, E.G. Finantu-Dinu, D.~Korzec, and J.~Engemann.
\newblock High-speed photographs of a dielectric barrier atmospheric pressure
  plasma jet.
\newblock \emph{{IEEE} Transactions on Plasma Science}, 33\penalty0
  (2):\penalty0 310--311, apr 2005.
\newblock \doi{10.1109/tps.2005.845377}.
\newblock URL \url{http://dx.doi.org/10.1109/tps.2005.845377}.

\bibitem[Sretenovi{\'{c}} et~al.(2011)Sretenovi{\'{c}}, Krsti{\'{c}},
  Kova{\v{c}}evi{\'{c}}, Obradovi{\'{c}}, and Kuraica]{Sretenovi__2011}
Goran~B. Sretenovi{\'{c}}, Ivan~B. Krsti{\'{c}}, Vesna~V.
  Kova{\v{c}}evi{\'{c}}, Bratislav~M. Obradovi{\'{c}}, and Milorad~M. Kuraica.
\newblock Spectroscopic measurement of electric field in atmospheric-pressure
  plasma jet operating in bullet mode.
\newblock \emph{Applied Physics Letters}, 99\penalty0 (16):\penalty0 161502,
  oct 2011.
\newblock \doi{10.1063/1.3653474}.
\newblock URL \url{http://dx.doi.org/10.1109/tps.2012.2219077}.

\bibitem[Naidis(2010)]{Naidis_2010}
G~V Naidis.
\newblock Modelling of streamer propagation in atmospheric-pressure helium
  plasma jets.
\newblock \emph{Journal of Physics D: Applied Physics}, 43\penalty0
  (40):\penalty0 402001, sep 2010.
\newblock \doi{10.1088/0022-3727/43/40/402001}.
\newblock URL \url{http://dx.doi.org/10.1063/1.3576940}.

\bibitem[Robert et~al.(2015)Robert, Darny, Dozias, Iseni, and
  Pouvesle]{Robert_2015}
E.~Robert, T.~Darny, S.~Dozias, S.~Iseni, and J.~M. Pouvesle.
\newblock New insights on the propagation of pulsed atmospheric plasma streams:
  From single jet to multi jet arrays.
\newblock \emph{Physics of Plasmas}, 22\penalty0 (12):\penalty0 122007, dec
  2015.
\newblock \doi{10.1063/1.4934655}.
\newblock URL \url{http://dx.doi.org/10.1063/1.4934655}.

\bibitem[Bourdon et~al.(2016)Bourdon, Darny, Pechereau, Pouvesle, Viegas,
  Is{\'{e}}ni, and Robert]{Bourdon_2016}
A~Bourdon, T~Darny, F~Pechereau, J-M Pouvesle, P~Viegas, S~Is{\'{e}}ni, and
  E.~Robert.
\newblock Numerical and experimental study of the dynamics of a$\mu$s helium
  plasma gun discharge with various amounts of {N\textsubscript{2}}admixture.
\newblock \emph{Plasma Sources Science and Technology}, 25\penalty0
  (3):\penalty0 035002, mar 2016.
\newblock \doi{10.1088/0963-0252/25/3/035002}.
\newblock URL \url{http://dx.doi.org/10.1088/0963-0252/25/3/035002}.

\bibitem[Darny(2016)]{Darny_2016}
Thibault Darny.
\newblock \emph{\'{E}tude de la production des esp\`{e}ces r\'{e}actives de
  l'oxyg\`{e}ne et de l'azote par d\'{e}charge Plasma Gun \`{a} pression
  atmosph\'{e}rique pour des applications biom\'{e}dicales}.
\newblock phdthesis, Universit\'{e} d'Orl\'{e}ans, June 2016.

\bibitem[Gaborit et~al.(2014{\natexlab{a}})Gaborit, Reuter, Iseni, and
  Duvillaret]{GaboritReuterIseniEtAl2014}
G.~Gaborit, S.~Reuter, S.~Iseni, and L.~Duvillaret.
\newblock {Cold Plasma Diagnostic Using Vectorial Electrooptic Probe}.
\newblock In \emph{{5\textsuperscript{th} International Conference on Plasma
  Medicine (ICPM5)}}, Nara, Japan, 18-23 May 2014{\natexlab{a}}.
\newblock \doi{10.13140/RG.2.2.12378.13768}.
\newblock URL \url{http://dx.doi.org/10.13140/RG.2.2.12378.13768}.

\bibitem[Gaborit et~al.(2014{\natexlab{b}})Gaborit, Jarrige, Lecoche, Dahdah,
  Duraz, Volat, and Duvillaret]{Gaborit_2014}
Gwenael Gaborit, Pierre Jarrige, Frederic Lecoche, Jean Dahdah, Eric Duraz,
  Christophe Volat, and Lionel Duvillaret.
\newblock Single shot and vectorial characterization of intense electric field
  in various environments with pigtailed electrooptic probe.
\newblock \emph{{IEEE} Transactions on Plasma Science}, 42\penalty0
  (5):\penalty0 1265--1273, may 2014{\natexlab{b}}.
\newblock \doi{10.1109/tps.2014.2301023}.
\newblock URL \url{http://dx.doi.org/10.1109/tps.2014.2301023}.

\bibitem[Darny et~al.(2017)Darny, Pouvesle, Puech, Douat, Dozias, and
  Robert]{Darny_2017}
T~Darny, J-M Pouvesle, V~Puech, C~Douat, S~Dozias, and Eric Robert.
\newblock Analysis of conductive target influence in plasma jet experiments
  through helium metastable and electric field measurements.
\newblock \emph{Plasma Sources Science and Technology}, 26\penalty0
  (4):\penalty0 045008, mar 2017.
\newblock \doi{10.1088/1361-6595/aa5b15}.
\newblock URL \url{http://dx.doi.org/10.1088/1361-6595/aa5b15}.

\bibitem[Norberg et~al.(2015)Norberg, Johnsen, and Kushner]{Norberg_2015}
Seth~A. Norberg, Eric Johnsen, and Mark~J. Kushner.
\newblock Helium atmospheric pressure plasma jets touching dielectric and metal
  surfaces.
\newblock \emph{Journal of Applied Physics}, 118\penalty0 (1):\penalty0 013301,
  jul 2015.
\newblock \doi{10.1063/1.4923345}.
\newblock URL \url{http://dx.doi.org/10.1063/1.4923345}.

\bibitem[Raizer(1991)]{9783642647604}
Yuri~P. Raizer.
\newblock \emph{Gas Discharge Physics}.
\newblock Springer-Verlag Berlin Heidelberg, 1991.
\newblock ISBN 364264760X.

\bibitem[Is\'{e}ni et~al.(2016{\natexlab{a}})Is\'{e}ni, Damany, Sretenovi\'{c},
  Kova\v{c}evi\'{c}, Krsti\'{c}, Dozias, Pouvesle, Kuraica, and
  Robert]{Iseni2016a}
S.~Is\'{e}ni, X.~Damany, G.~Sretenovi\'{c}, V.~Kova\v{c}evi\'{c},
  I.~Krsti\'{c}, S.~Dozias, J.-M. Pouvesle, M.~Kuraica, and E.~Robert.
\newblock Electric field and discharge properties of single and multiple
  arrangement of pulsed atmospheric plasma streams.
\newblock In \emph{{28\textsuperscript{th} Summer School and International
  Symposium on the Plysics of Ionized Gases (SPIG2016)}}, Belgrade, Serbia,
  29\textsuperscript{th} Aug. - 2\textsuperscript{nd} Sep. 2016{\natexlab{a}}.
\newblock ISBN 978-86-84539-14-6.

\bibitem[Is\'{e}ni et~al.(2016{\natexlab{b}})Is\'{e}ni, Damany, Darny, Douat,
  Dozias, Pouvesle, and Robert]{Iseni2016}
S.~Is\'{e}ni, X.~Damany, T.~Darny, C.~Douat, S.~Dozias, J.-M. Pouvesle, and
  E.~Robert.
\newblock Electric field characterization of plasma gun and multi-jet plasma
  arrays.
\newblock In \emph{{6\textsuperscript{th} International Conference on Plasma
  Medicine (ICPM6)}}, Bratislava, Slovakia, 04-09 September 2016{\natexlab{b}}.

\bibitem[Wild et~al.(2013)Wild, Gerling, Bussiahn, Weltmann, and
  Stollenwerk]{Wild_2013}
R~Wild, T~Gerling, R~Bussiahn, K-D Weltmann, and L~Stollenwerk.
\newblock Phase-resolved measurement of electric charge deposited by an
  atmospheric pressure plasma jet on a dielectric surface.
\newblock \emph{Journal of Physics D: Applied Physics}, 47\penalty0
  (4):\penalty0 042001, dec 2013.
\newblock \doi{10.1088/0022-3727/47/4/042001}.
\newblock URL \url{http://dx.doi.org/10.1088/0022-3727/47/4/042001}.

\bibitem[Slikboer et~al.(2017)Slikboer, Garcia-Caurel, Guaitella, and
  Sobota]{Slikboer_2017}
Elmar Slikboer, Enric Garcia-Caurel, Olivier Guaitella, and Ana Sobota.
\newblock Charge transfer to a dielectric target by guided ionization waves
  using electric field measurements.
\newblock \emph{Plasma Sources Science and Technology}, 26\penalty0
  (3):\penalty0 035002, feb 2017.
\newblock \doi{10.1088/1361-6595/aa53fe}.
\newblock URL \url{http://dx.doi.org/10.1088/1361-6595/aa53fe}.

\bibitem[Xiong and Kushner(2012)]{Xiong_2012}
Zhongmin Xiong and Mark~J Kushner.
\newblock Atmospheric pressure ionization waves propagating through a flexible
  high aspect ratio capillary channel and impinging upon a target.
\newblock \emph{Plasma Sources Science and Technology}, 21\penalty0
  (3):\penalty0 034001, apr 2012.
\newblock \doi{10.1088/0963-0252/21/3/034001}.
\newblock URL \url{http://dx.doi.org/10.1088/0963-0252/21/3/034001}.

\bibitem[Naidis(2011)]{Naidis_2011}
G.~V. Naidis.
\newblock Simulation of streamers propagating along helium jets in ambient air:
  Polarity-induced effects.
\newblock \emph{Applied Physics Letters}, 98\penalty0 (14):\penalty0 141501,
  apr 2011.
\newblock \doi{10.1063/1.3576940}.
\newblock URL \url{http://dx.doi.org/10.1063/1.3576940}.

\bibitem[Xiong et~al.(2013)Xiong, Robert, Sarron, Pouvesle, and
  Kushner]{Xiong_2013}
Zhongmin Xiong, Eric Robert, Vanessa Sarron, Jean-Michel Pouvesle, and Mark~J
  Kushner.
\newblock Atmospheric-pressure plasma transfer across dielectric channels and
  tubes.
\newblock \emph{Journal of Physics D: Applied Physics}, 46\penalty0
  (15):\penalty0 155203, mar 2013.
\newblock \doi{10.1088/0022-3727/46/15/155203}.
\newblock URL \url{http://dx.doi.org/10.1088/0022-3727/46/15/155203}.

\bibitem[Sretenovi{\'c} et~al.(2014)Sretenovi{\'c}, Krsti{\'c},
  Kova{\v{c}}evi{\'c}, Obradovi{\'c}, and Kuraica]{Sretenovic2014}
Goran~B Sretenovi{\'c}, Ivan~B Krsti{\'c}, Vesna~V Kova{\v{c}}evi{\'c},
  Bratislav~M Obradovi{\'c}, and Milorad~M Kuraica.
\newblock Spatio-temporally resolved electric field measurements in helium
  plasma jet.
\newblock \emph{Journal of Physics D: Applied Physics}, 47\penalty0
  (10):\penalty0 102001, 2014.
\newblock \doi{10.1088/0022-3727/47/10/102001}.
\newblock URL \url{http://dx.doi.org/10.1088/0022-3727/47/10/102001}.

\bibitem[van~der Walt et~al.(2011)van~der Walt, Colbert, and
  Varoquaux]{van_der_Walt_2011}
St{\'{e}}fan van~der Walt, S~Chris Colbert, and Ga{\"{e}}̈l Varoquaux.
\newblock The {NumPy} array: A structure for efficient numerical computation.
\newblock \emph{Computing in Science {\&} Engineering}, 13\penalty0
  (2):\penalty0 22--30, mar 2011.
\newblock \doi{10.1109/mcse.2011.37}.
\newblock URL \url{http://dx.doi.org/10.1109/mcse.2011.37}.

\bibitem[Hunter(2007)]{Hunter_2007}
John~D. Hunter.
\newblock Matplotlib: A {2D} graphics environment.
\newblock \emph{Computing in Science {\&} Engineering}, 9\penalty0
  (3):\penalty0 90--95, 2007.
\newblock \doi{10.1109/mcse.2007.55}.
\newblock URL \url{http://dx.doi.org/10.1109/mcse.2007.55}.

\end{thebibliography}
\end{document}